\documentclass[11pt,preprint]{aastex}

\def\etal{ et~al.\rm}

\def\ergss{ergs s$^{-1}$\,}
\def\chan{{\it Chandra\,}}
\def\rosat{{\it ROSAT}}
\def\eins{{\it Einstein}}
\def\wavd{{\it wavdetect}\,}
\def\xmm{{\it XMM-Newton}}
\def\lx{L$_{X}$}

\begin{document}

\title{Revealing the Supernova Remnant Population of M33 with Chandra}

\author{Parviz Ghavamian\altaffilmark{1}, William P. Blair\altaffilmark{1}, Knox S. Long\altaffilmark{2},
Manami Sasaki\altaffilmark{3}, Terrance J. Gaetz\altaffilmark{3} and Paul P. Plucinsky\altaffilmark{3} }

\altaffiltext{1}{Department of Physics and Astronomy, Johns Hopkins University, 3400 North
Charles Street, Baltimore, MD, 21218-2686; parviz@pha.jhu.edu, wpb@pha.jhu.edu}
\altaffiltext{2}{Space Telescope Science Institute, 3700 San Martin Drive, Baltimore, MD, 21218; long@stsci.edu}
\altaffiltext{3}{Harvard-Smithsonian Center for Astrophysics, 60 Garden Street, Cambridge, MA, 02138; sasaki@cfa.harvard.edu,
gaetz@cfa.harvard.edu, plucinsky@cfa.harvard.edu  }

\begin{abstract}

We present results of a search for supernova remnants (SNRs) in archival \chan\, images
of M33.  We have identified X-ray SNRs by comparing the list of \chan\, X-ray sources in M33
with tabulations of SNR candidates identified from (1) elevated
[S II]/H$\alpha$ ratios in the optical, and (2) radio spectral indices.  In
addition, we have searched for optical counterparts to soft sources in the \chan\,
images and X-ray SNR candidates identified in the \xmm\, survey of M33.  Of the 98 optically known
SNRs in M33, 22 have been detected at $>$3$\sigma$ level in the soft band (0.35$-$1.1 keV).  At least four
of these SNR candidates are spatially extended based on a comparison of the data
to simulated images of point sources.  Aside from the optically matching SNRs, we have found one soft X-ray
source in M33 which exhibits no optical emission
and is coincident with a known radio source.  The radio spectral index of this source is consistent with
particle acceleration in shocks, leading us to suggest that it is
a non-radiative SNR.  We have also found new optical counterparts to two soft X-ray SNRs in M33.  These
counterparts exhibit enhanced [S~II]/H$\alpha$ ratios characteristic of radiative shocks.  Pending
confirmation from optical spectroscopy, the identification of these two optical counterparts increases the total
number of known optically emitting SNRs in M33 to 100.  This brings the total number of identified
SNRs with X-ray counterparts, including those exclusively detected by the \xmm\, survey of M33, to 37 SNRs.
We find that while there are a similar number
of confirmed X-ray SNRs in M33 and the LMC with X-ray luminosities in excess of 10$^{35}$
\ergss, nearly 40\% of the LMC SNRs are brighter than 10$^{36}$ \ergss, while only 13\% of
the M33 sample exceed this luminosity.  Including X-ray SNR candidates from the \xmm\, survey
(objects lacking optical counterparts) increases the fraction of M33 SNRs brighter than 10$^{36}$ \ergss\,
to 22\%, still only half the LMC fraction.  The differences in luminosity distributions
cannot be fully explained by uncertainty in spectral model parameters, and is not fully accounted 
for by abundance differences between the galaxies.  

\end{abstract}


\section{INTRODUCTION}

M31, M33 and the Milky Way are the dominant galaxies of the
Local Group and are the nearest normal galaxies that can 
be studied in detail.  In particular, the proximity of M33 (795$\pm$75 kpc;
van den Bergh 1991), its low inclination ($<$ 55$^{\circ}$; Zaritsky
\etal\, 1989) and modest foreground extinction (N$_{H}$(Gal)\,$\lesssim\,$6$\times$10$^{20}$
cm$^{-2}$; Stark \etal\, 1992) have made this stellar system one of the best 
studied galaxies.  Classified as a late-type ScII-III spiral, M33
is intermediate between the more massive early-type spirals such as the
Milky Way and M31 and the dwarf irregular galaxies such as the Magellanic
Clouds.  M33 exhibits a large number of OB associations, H~II regions and
supershells (Boulesteix 1974, Viallefond \etal\, 1986), indicating that it is host to a large
number of active star forming regions.  

The Local Group galaxies are host to a large number of SNRs.  These objects
are fundamental to our understanding of the interstellar
medium (ISM) and to changes in the composition of galaxies over time.  They are probes of
and a major energy input source to the ISM.  The distribution of supernovae, and hence SNRs,
determines how much of the ISM is hot (T\,$\sim$\,10$^{6}$~K); therefore, emission from
SNRs is intimately connected with the soft X-ray background in nearby galaxies.  

Multiwavelength surveys of M33 have been especially useful for
understanding the global properties of SNRs in that galaxy.
These surveys are highly useful tools for probing the global properties of
nearby galaxies, but they are also subject to limitations.
Variations in detector resolution and sensitivity across different wavelength bands result
in varying degrees of completeness in radio, optical and X-ray source catalogues.  In addition,
physical processes such as absorption can affect some wavebands (such as the soft X-ray band)
more than others (such as the optical band).  This is particularly true of 
soft X-ray sources such as SNRs, which are easily rendered undetectable even in moderately
absorbed regions.

Altogether, the most successful searches for SNRs have been
performed in the optical for galaxies such as the LMC and SMC (Mathewson \& Clarke 1973;
Mathewson \etal\, 1983, 1984, 1985), M31 (Blair, Kirshner \& Chevalier 1982), M33 (Sabbadin
1977, 1979; D'Odorico, Dopita \& Benvenuti 1980; Blair \& Kirshner 1985; Viallefond \etal\, 1986; Long 
\etal\, 1990; Smith \etal\, 1993, Gordon \etal\, 1998; hereafter GKL98), the Sculptor Group galaxies NGC 300 and 7793 
(Blair \& Long 1997), M83 (Blair \& Long 2004) and the nearby spirals NGC 2403, 5204, 5585, 6946, M81 and M101 (Matonick
\etal\, 1997, Matonick \& Fesen 1997).
The optical identification technique consists of dividing continuum-subtracted, narrowband [S~II] images of
galaxies by narrowband H$\alpha$ images and then searching the ratio image for features with elevated [S~II]/H$\alpha$
ratios.  There is generally a strong separation in [S~II]/H$\alpha$ ratio between H~II regions ([S~II]/H$\alpha\,\lesssim\,$
0.1) and SNRs ([S~II]/H$\alpha$\,$\gtrsim$\,0.4) (Mathewson \& Clarke 1973).
The ratio differences are due to the fact that photoionization keeps S ionized
beyond S$^{+}$, resulting in weak [S~II] emission.  On the other hand, radiative recombination in SNRs produces 
a wide range of temperatures and ionization states so that a zone exists behind the shock where a larger fraction of S resides in S$^{+}$ and
[S~II] emission is strong.

M33 has been surveyed by each successive X-ray mission up to the present day.  These include surveys by
\eins\, (Long \etal\, 1981; Trinchieri \etal\, 1988), \rosat\, (Schulman \& Bregman 1995, Long \etal\, 1996,
Haberl \& Pietsch 2001) and \xmm\, (Pietsch \etal\, 2003, 2004).  The \eins\, High Resolution Imager (HRI)
and Imaging Proportional Counter (IPC) surveys
showed that the X-ray appearance of M33 is dominated by a hard nuclear source.   The luminosity of
this object (M33 X-8) was found to be $\sim\,$10$^{39}$ \ergss\, (Long \etal\, 1996), a result
later confirmed by \chan\, observations (Dubus \& Rutledge 2002).  This makes M33 X-8 the brightest X-ray source in the Local
Group.  The 14 other unresolved sources found in the \eins\, survey (\lx(0.1$-$2.4 keV)\,$\gtrsim$10$^{37}$ \ergss,
Long \etal\, 1996) were classified as X-ray binaries.  Subsequent surveys with the HRI
and Position Sensitive Proportional Counter (PSPC) on \rosat\, revealed faint diffuse emission
within 10\arcmin\, of the nucleus and increased the total number of detected X-ray
sources to 184.  The \rosat\, survey depth was \lx(0.1$-$2.4 keV)\,$\approx$\,10$^{36}$ \ergss\, (Long 
\etal\, 1996), resulting in the detection of 12 of the 98 optically identified SNRs
in M33.  The most recent X-ray survey of M33 was performed with \xmm\, 
and covered the full D$_{25}$ ellipse with uniform sensitivity down to 
a luminosity \lx(0.5$-$10 keV)\,$\approx\,$10$^{35}$ \ergss\, (Pietsch \etal\, 2003, 2004).  That
survey brought the total number of sources detected in M33 to 408, including 28 sources matching
optical SNRs from the catalogue of GKL98.  

Here we present results of a search for supernova
remnants (SNRs) in archival \chan\, images of M33, utilizing positive coincidence with
optically known SNRs and hardness ratios as discriminants.  We focus our analysis on the
SNR population, and do not attempt to characterize the properties of the global distribution of
X-ray sources in this galaxy (a broader discussion is provided by Grimm \etal\, 2005, ApJ, submitted.).
There were 98 known optical SNRs in M33 when we began this work, of which 78 lay in the field of view
of the \chan\, archival images.  Of these, only 26 were located within a 10\arcmin\, diameter circle where the
\chan\, point spread function is characterized by a 50\% encircled
energy (E\,=\,1.49 keV, \chan\, Proposers' Guide, v.6) of 2\farcs5 or less.
We have found X-ray counterparts to at least 22 optically known SNRs in the archival \chan\,
images of M33. We have also found two previously unidentified optical SNRs in M33 by comparing \chan\, and \xmm\,
source lists with narrowband optical images.  

\section{Observations and Data Reduction}

\subsection{\chan\, Images}

The \chan\, archival data used in our analysis were acquired during Cycle 1 (ObsID 786 in
ACIS-S imaging mode, ObsID 1730 in ACIS-I imaging mode) and Cycle 2 (ObsID 2023 in ACIS-I
imaging mode).  The Cycle 1 observations targeted the bright
nuclear source of M33 (aimpoints at $\alpha_{J2000}$\,=\,01$^{\rm h}$33$^{\rm m}$50.8$^{\rm s}$,
$\delta_{J2000}\,=\,$30$^{\circ}$39\arcmin\,36\farcs6), while the Cycle 2 pointing was centered on NGC 604, the giant starburst H~II region
along the northern spiral arm (aimpoint at $\alpha_{J2000}$\,=\,01$^{\rm h}$34$^{\rm m}$32.9$^{\rm s}$,
$\delta_{J2000}\,=\,$30$^{\circ}$47\arcmin\,04\farcs0).  The \chan\, imaging fields are marked on a continuum-subtracted
H$\alpha$ image of M33 in Fig.~1.  The combined field of view of the \chan\, images 
covers approximately half of the region determined by the D$_{25}$ isophote of M33 (Tully 1988).

Using CIAO version 3.0.2, we applied the Chandra X-ray Center (CXC) CTI correction to the level 1 events files
and screened the data to remove time intervals with background rates $\geq$4$\sigma$ above the median 
level and restrict the energy range of the
resulting datasets to 0.35$-$8 keV.  We then applied the CIAO destreaking algorithm to remove the
streak pattern from the chip.  Finally, we used the CXC aspect offset tool to correct the WCS information
for each events file.  The resulting exposure times for
the ObsID 786, 1730 and 2023 datasets were 46.3 ks, 49.4 ks and 88.8 ks, respectively.

\subsection{Narrowband Optical Imagery}

To search for optical counterparts to the X-ray sources in the \chan\, data we retrieved KPNO 4m Mosaic
images of M33 from the NOAO data archive.  The optical data include narrowband imagery in H$\alpha$, 
[S~II] and [O~III] and cover most of the D$_{25}$ isophote from the Local Group Survey of Massey \etal\, 
(2002)\footnote{This research draws upon data provided by P. Massey as distributed by the NOAO Science
Archive.  NOAO is operated by the Association of Universities for Research in Astronomy (AURA), Inc.
under a cooperative agreement with the National Science Foundation. }.  The images we used from the
archive were the final combined frames (stack of 5 dithered frames) and were overscan and bias subtracted,
flat fielded and corrected for bad pixels.

To remove the continuum emission from the narrowband images we first subtracted a constant pedestal value from 
the H$\alpha$, [S~II] and R-band images of M33 to
bring all the frames to a zero level background.  We then scaled and subtracted the R-band image from the H$\alpha$ and [S~II]
frames to remove stellar continuum from the narrowband images.  Similarly, we used the B-band image to subtract
the continuum emission from the [O~III] image. We then divided the continuum-subtracted
[S~II] and H$\alpha$ frames to produce a [S~II]/H$\alpha$ ratio image.  To ensure the reliability
of the [S~II]/H$\alpha$ ratios measured from the NOAO images, we compared values of the ratio
at positions of known optical SNRs to the ratios
listed in Table~3 of GKL98.  Our calculated [S~II]/H$\alpha$ ratios agreed with the tabulated values of
GKL98 to within 20\%, and all known SNRs were readily distinguishable in the ratio image.

\section{X-ray Source Detection}

We used the CIAO routine \wavd, which detects sources by
convolving the pixels with ``Mexican Hat'' wavelet functions for source detection (Freeman \etal\, 2002).
Optical surveys of M33 (Long et al. 1990, GKL98) have shown that the SNRs in this galaxy
span a wide range of radii, from $\sim$ 1\arcsec\, ($\sim$ 4 pc)     
up to $\sim$ 15\arcsec\, ($\sim$ 60 pc).  Accordingly, we optimized our search
for spatially extended X-ray emission from SNRs by conducting our \wavd\, runs
on wavelet scale sizes up to 64\arcsec.  The wavelet sizes utilized during our runs
were (0\farcs5, 1\arcsec, 2\arcsec, 4\arcsec, 8\arcsec, 16\arcsec,
32\arcsec, 64\arcsec).  The largest threshold significance for output source 
lists was set to 10$^{-6}$.  

Before searching for sources in the data we used 1 keV exposure maps to correct
the count rates of each image.  We input the exposure map into each \wavd\, run
to avoid detection of spurious sources from such features as charge transfer streaks.  We 
filtered each events file to create images in each of the following bands: 
0.35$-$1.1~keV (soft), 1.1$-$2.6~keV (medium), 2.6$-$8.0~keV (hard)
and 0.35$-$8.0~keV (broad).  We performed the source detection on each filtered
image separately.  Since the imaging fields of the three datasets overlap,
we scanned the source detection output for multiply detected sources.  If a source appeared
in more than one observation with similar S/N in both images, 
we retained the detection parameters from the observation with the
best imaging quality (i.e., the one where the source was closest to the ACIS aimpoint).  
On the other hand, exceptions were made if a multiply detected source exhibited significantly higher
S/N in one image than another, in which case we retained detection parameters from the 
image with the stronger detection.  The number of unique sources detected in the broad band
images of ObsID 786, 1730 and 2023 was 166 (207) at the 3$\sigma$ (2$\sigma$) level.  

\section{Cross-Correlation of \chan\, Sources with Optical and Radio Catalogues}

As is evident from the optical
images of M33 (GKL98), many SNRs in this galaxy are expanding into inhomogeneous
media, with many exhibiting distorted morphologies.
The likely mixture of radiative and non-radiative shocks in these SNRs can produce wide
variations in X-ray and optical brightness along their shells.  Depending on
whether the shocks are radiative, non-radiative or a combination, portions of the
shell may be detected solely in the optical, solely in X-rays, or in both bands.  
Furthermore, unlike many SNRs studied in galaxies outside the Local Group, the M33 SNRs 
are often large enough ($\gtrsim$3\arcsec across, or 12 pc) to be spatially 
resolved in optical and \chan\, X-ray images.  The net effect is that choosing a 
fixed matching radius during the cross-correlation can cause real
matches to be missed.  To avoid this problem, we perform the cross-correlation in 
two steps.  First, we obtained a culled list of sources in all three bands that match
the coordinates of GKL98 SNRs to within a generous 
coincidence radius of 20\arcsec.  
We then generated a final list of X-ray counterparts to the GKL98 SNRs by 
visually inspecting each matching object.  We blinked between aligned X-ray and H$\alpha$ 
images of each source to confirm that the X-ray source lay within the optically measured
diameter of the SNR.  We also looked for secondary signs of a match, such as  
evidence of extended morphology.  However, this feature is not a strong discriminant,
since the interaction of SNRs with compact ISM clouds can produce localized regions of enhanced X-ray emission
unresolved by \chan.  Although we have taken advantage of overlaps
between the three \chan\, observations wherever possible to select SNR candidates with the best imaging quality 
and highest count levels, the final group of candidates are invariably affected by strong variations
in the size of the \chan\, PSF between datasets and within each individual observation.
This is the most significant complicating factor in distinguishing point sources 
from extended sources.

In the majority of cases the X-ray emission from SNRs is dominated by thermal 
emission (lines + bremsstrahlung continuum) from shocks in ISM and ejecta
material.  The thermal emission is typically soft, peaking below 1 keV. Therefore, we
concentrated our search for SNRs on the soft band images (0.35$-$1.1~keV) of M33.

\section{Supernova Remnants Detected with \chan}

\subsection{Optical Matches}

Of the detected sources in ObsID datasets 786, 1730 and 2023, we find that 22 match
SNRs from the optical catalogue, all at $\geq$3$\sigma$ in the S band (Table~1).
Images of the matching sources are shown in Figures \ref{m33_thumb1}, \ref{m33_thumb2}, \ref{m33_thumb3}\,
and \ref{m33_thumb4}.  

Before we attempted to interpret the results of our matches, we first
calculated the expected number of random coincidences between the two source lists.
As shown in Figure~1, the \chan\, observations of M33 do not cover the entire spiral.  
Although the \chan\, images cover a smaller fraction of the M33 spiral than the earlier \rosat\,
images (Long \etal\, 1996, Haberl \& Pietsch 2001), the higher sensitivity and better spatial resolution 
of \chan\, have resulted in a greater number of X-ray detections of optically identified SNRs.  Of the
98 SNRs catalogued by GKL98, 78 lie within the total field of view of the three
\chan\, observations.  Since M33 fills the entire X-ray field of view, we can expect to find
both background objects (AGN) and M33 sources on each ACIS chip.  Assuming these objects
are distributed randomly across the field, the 
number of chance coincidences between the 166 X-ray sources and the 78 optical SNRs in the \chan\,
field is $<$\,3
for a matching radius of 15\arcsec.   This may be an underestimate, given that X-ray sources
intrinsic to M33 are not randomly distributed.  However, the culling of individual matching sources
by spectral hardness (and visual inspection as a secondary indicator) reduces the likelihood 
that chance coincidences are retained in the final list of matches.

SNRs are extended objects. To check whether spatial extent could provide an additional criterion
for selecting SNRs from the X-ray sample, we visually compared each SNR candidate with a simulated X-ray image
of a point source containing the same number of counts and located at the same off axis angle as the SNR candidate.
We performed these simulations at 1.5 keV using the CXC applications ChaRT and MARX 4.0.8.
SNR candidates which clearly appeared to be larger than their corresponding simulated sources were labeled as
extended (Table~1).  Since our simulated point source images did not include background emission, they
were of limited usefulness in identifying extended SNRs that were faint and/or located far from the
imaging axis.  The inability to separate background fluctuations from real extended
emission made 10 of the SNR candidates unsuitable for direct visual comparison with point source
images.  However, 4 sources showed clear extended morphology, while 3 others showed marginal evidence
of extended emission and 5 showed morphologies consistent with unresolved (point source) emission.
Note that since the X-ray emission from some sources may originate in localized
regions smaller than the \chan\, imaging resolution, a test for spatial extent can only confirm
the SNR nature of a source, not disprove it. 

Aside from extended morphology, SNRs are also expected to show temporally steady fluxes. 
We searched for time variability in our SNR candidates by converting \xmm\, count rates of optically matching
candidates to expected \chan\, count rates, then comparing these rates with the \chan\, values from
Table~2.  We performed the conversion using the best fit parameters for each source from Table~3 (see discussion below).
Of the 16 SNR candidates detected by both \xmm\, and \chan, all but 2 exhibit fluxes consistent to within
20\%.  The 0.35$-$10 keV \chan\, count rates predicted for GKL98 SNRs 28 and 29 by the \xmm\, data are nearly twice
the values obtained in Table~2.  The reason for the discrepancy is unclear, but may arise from the inclusion
of nearby soft diffuse emission in the \xmm\, spectral extraction regions of Pietsch \etal\, (2004).  On
the other hand, it is also possible that the match between these two X-ray sources and 
their optical counterparts is the result of a random coincidence.  In the case of SNR 28, at least,
this possibility is lessened by the marginally extended morphology of its X-ray counterpart.  However,
we have no such assurance for the SNR 29 counterpart, which shows no obvious extent.  Therefore, we
proceed with the remaining analysis of our paper with the caveat that two of the most luminous
soft X-ray sources in our sample may not be SNRs.

\subsection{Properties of X-ray Sources Matching Optical SNRs}

During the extraction of X-ray spectra from SNR candidates we were confronted with the
problem of choosing aperture regions large enough to accommodate both the extended morphologies of the 
sources and significant variations in the size of the \chan\, PSF across each field.  A lower limit
on the aperture size for a given source is obtained by assuming it is pointlike and
simply setting the aperture radius equal
to the size of the \chan\, PSF at that location.  An upper limit can be obtained by using 
both the PSF size and the (admittedly rough) optically measured sizes for the known optical SNR.
Empirically we found that a source extraction radius R$_{extr}$\,=\,R$_{PSF}$\,+\,1.5\,R$_{opt}$
was large enough to include all the S band emission from the SNR candidates.  We set
R$_{PSF}$ to the value enclosing 95\% of the counts at 1.5 keV (R$_{PSF}$ is a function of
off-axis angle, see Figure~4.12 in the \chan\, Proposers' Guide v6).  The extraction regions used in our
SNR analysis are marked on Figures \ref{m33_thumb1}\,-\,\ref{m33_thumb4}, and range from approximately
4\arcsec\, to 21\arcsec\, in radius.
We estimated the local X-ray background of each object using an annulus centered on the source, with
inner radius set to 2\,R$_{extr}$ and outer radius to 3\,R$_{extr}$.
Known X-ray sources lying within a given annulus were excluded from the background spectral
extraction, as were chip edges and diffuse emission from NGC 604.  For all the remaining sources
in the \chan\, observations we set the size of the extraction apertures to R$_{PSF}$.

Source counts were extracted in soft (S), medium (M) and hard (H) bands defined
above and used to compute the hardness ratios HR1\,$\equiv\,$(M\,$-$\,S)/(M\,+\,S) and HR2\,$\equiv$\,(H\,$-$\,M)
/(H\,+\,M) (Table~2).  In many cases the raw object and/or background spectra exhibited low count levels (N$\lesssim$ 20),
requiring the usage of Poisson statistics in defining the errors (rather than the commonly used Gaussian
statistics).  In cases where N\,$<$\,10 we utilized the Gehrels (1986) approximation to compute
the 1$\sigma$ errors on the count levels.  This results in an asymmetric distribution with the error bars on
the upper limit larger than on the lower limit.  In these cases we set the lower error bar equal to the upper
error bar before computing HR1 and HR2 and propagating the errors.

By this definition, soft sources such as SNRs and active stars tend to exhibit negative hardness ratios,
while hard sources such as AGN and X-ray binaries tend to exhibit positive hardness ratios.  This is generally a good
discriminant for separating thermal sources (soft) from non-thermal sources (hard).  However, 
among the X-ray sources lacking known optical counterparts there are also a number of objects exhibiting mixed hardness ratios.  
Determining the physical origin of these sources is more difficult, since they may exhibit a mixture of
thermal and non-thermal emission but are too faint for conclusive X-ray spectral fitting.

From Figure \ref{m33_hardratios}\, and Table~2 it is clear that the background-subtracted spectra of the SNR candidates exhibit
hardness ratios consistent with soft emission (HR1\,$\leq$\,0, HR2\,$\leq$\,0).  To within the errors this is consistent with thermal
emission from shocked plasma.  The spectra of many SNR candidates exhibit
negative counts in the H band after background subtraction, consistent with zero net counts.  In these
cases we report HR2\,=\,$-$1.  Some candidates, such as the counterparts to optical SNRs 37, 54, 62 and 85 even 
exhibit zero net counts in the M band, giving HR1\,=\,$-$1.  

\begin{figure}
\plotone{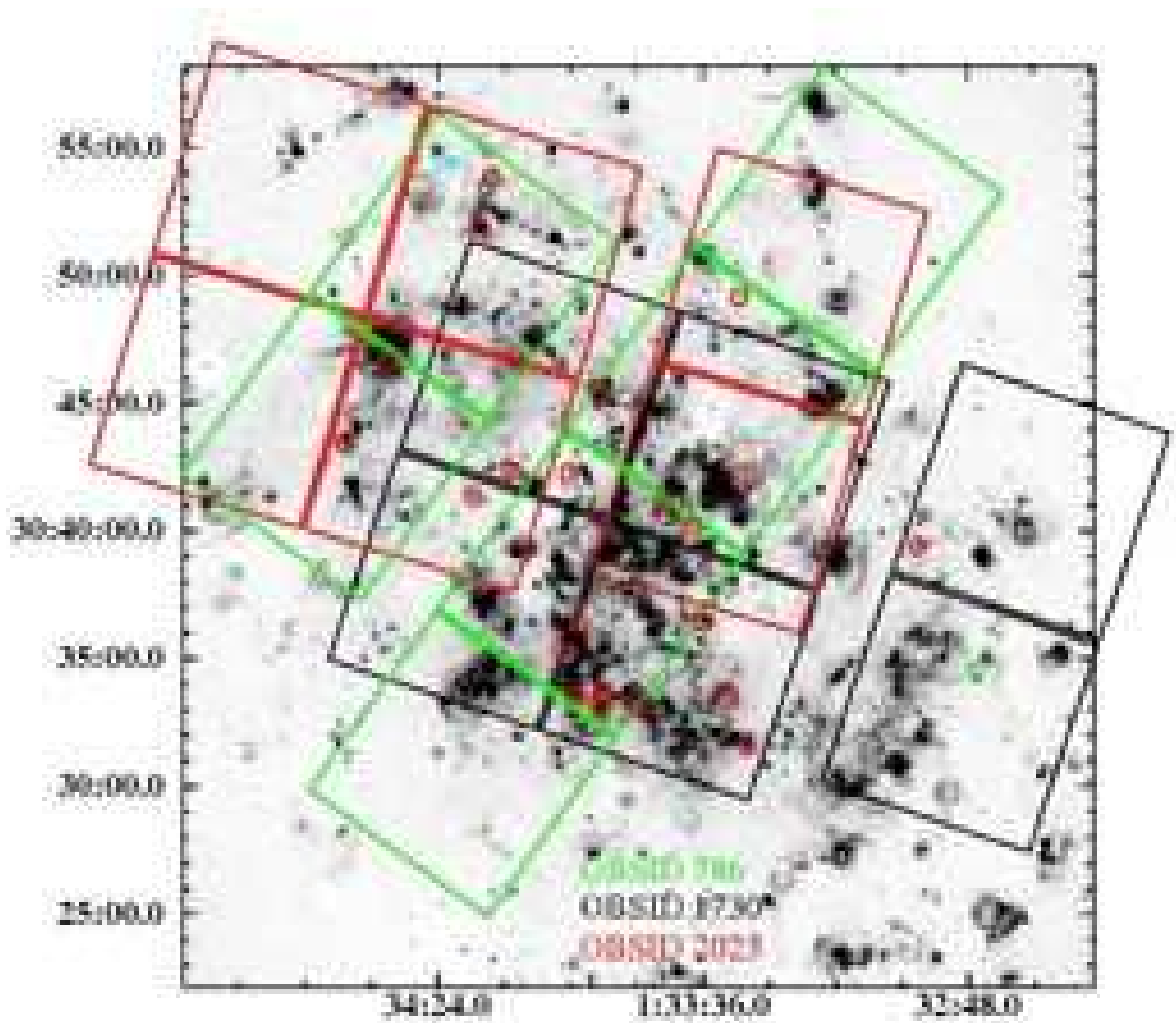}
\caption{
Continuum-subtracted H$\alpha$ image of M33 from the Local Group Survey (Massey \etal\, 2002), shown
with the ACIS detector footprints from the three archival \chan\, images.  The red circles mark the
positions of the 22 optically identified SNRs from the catalogue of GKL98 exhibiting
X-ray counterparts in the \chan\, images.  The cyan box marks the position of
a newly discovered optical SNR matching a soft X-ray source in the \chan\, and \xmm\, data,
while the green box marks the position of a new optical SNR matching a soft X-ray
source seen exclusively in the \xmm\, observations (Pietsch \etal\, 2004).  Details on
these two new optical SNRs are presented in Section 7.   A full resolution
figure can be found at http://fuse.pha.jhu.edu/\~{}parviz/papers/m33/. }
\label{m33_halpha}
\end{figure}

\begin{figure}
\figurenum{2}
\epsscale{0.7}
\plotone{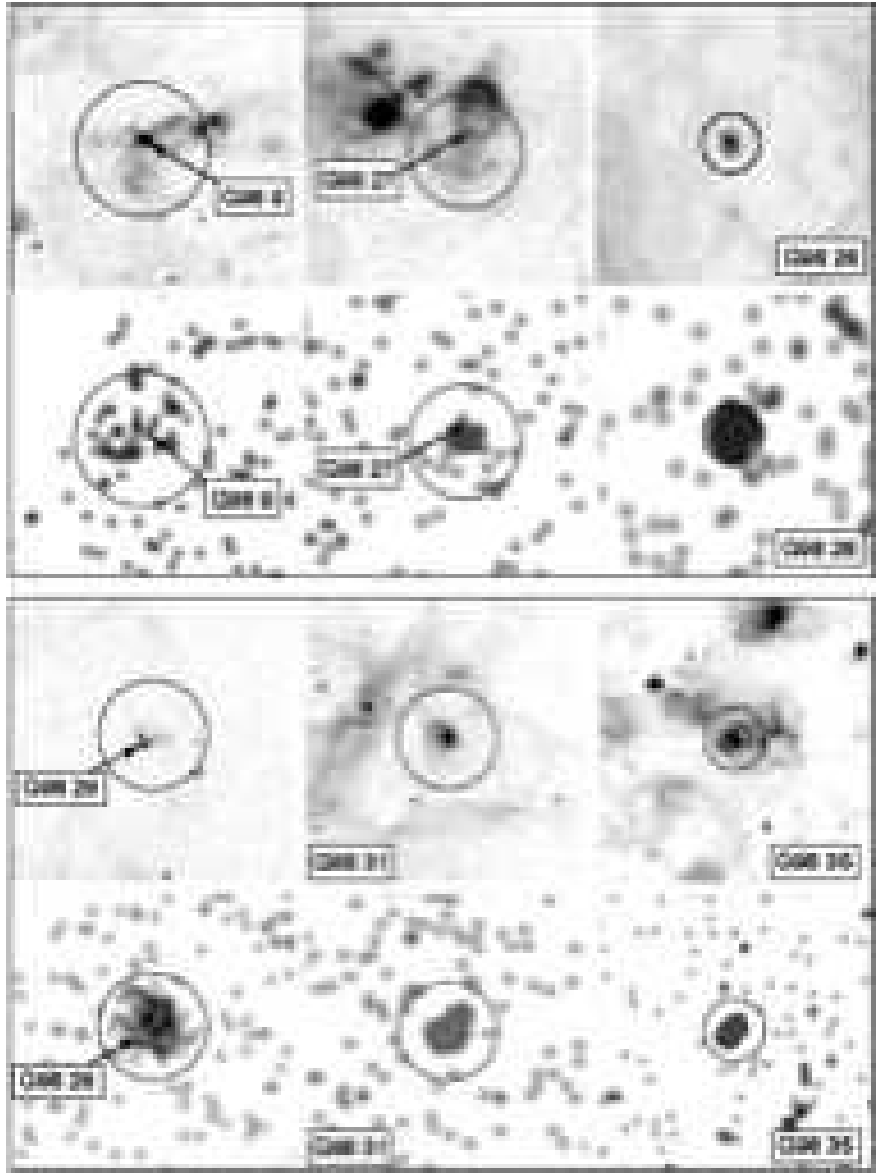}
\caption{
Closeup views of SNRs in M33 identified
in the optical by Gordon et al. (1998) and exhibiting X-ray counterparts in the {\it Chandra}
data.  The top frame in each sequence is a continuum-subtracted KPNO Mosaic image of the
SNR in H$\alpha$ (Massey \etal\, 2002), with the GKL98 catalogue number listed.  The bottom frame
shows the \chan\, counterpart of each SNR in the soft band (0.35$-$1.1 keV), smoothed with
a 3 pixel Gaussian.  Circles mark the
apertures used to extract X-ray spectra of each SNR. The arrows in both images point to 
the location of the [S~II]/H$\alpha$ enhancement measured by GKL98.  The thumbnail for
SNR 28 from the GKL98 catalogue is 1\farcm0 on the side, while the remaining thumbnails
are each 1\farcm5 on the side.  A full resolution
figure can be found at http://fuse.pha.jhu.edu/\~{}parviz/papers/m33/.}
\epsscale{1.0}
\label{m33_thumb1}
\end{figure}

\begin{figure}
\figurenum{3}
\epsscale{0.7}
\plotone{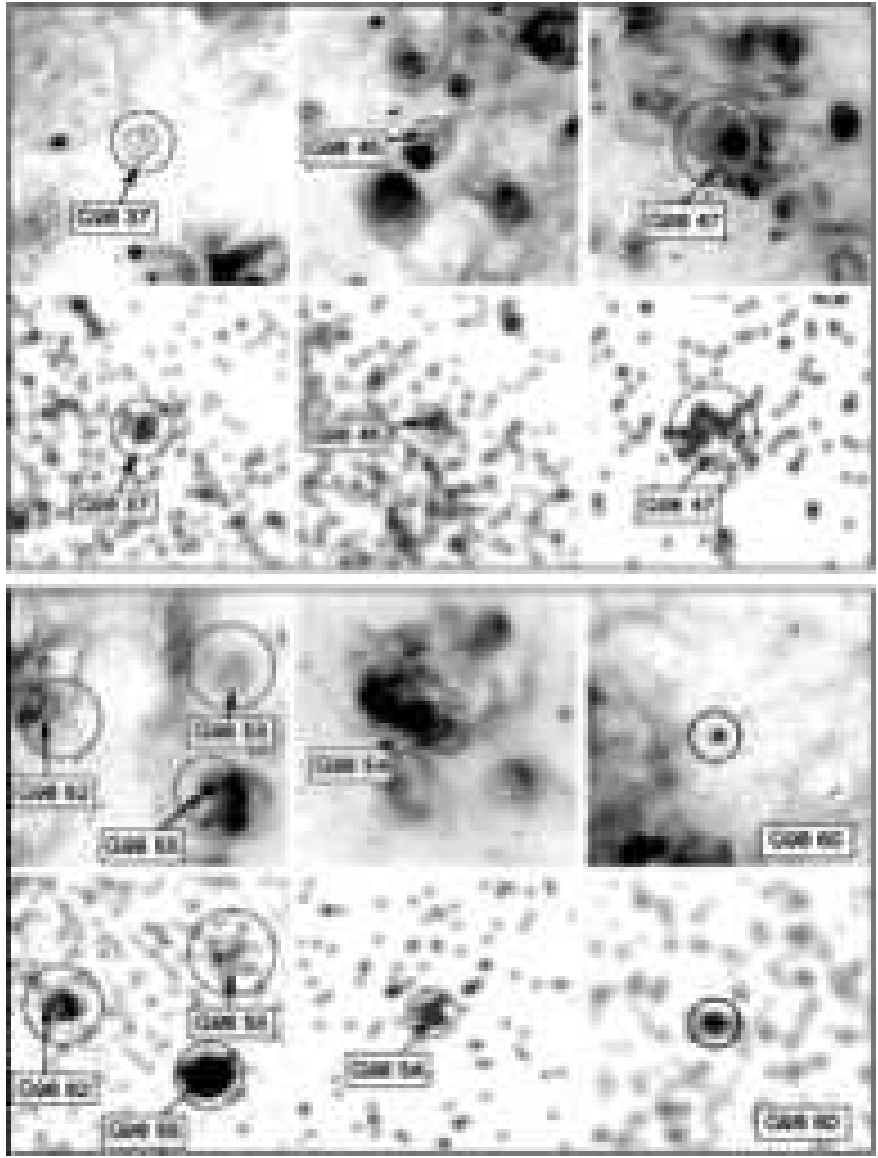}
\caption{Same as Figure \ref{m33_thumb1}.  The thumbnail for SNR 60 from the GKL98
catalogue is 1\farcm0 on the side, while the remaining thumbnails
are each 1\farcm5 on the side.  A full resolution
figure can be found at http://fuse.pha.jhu.edu/\~{}parviz/papers/m33/.}
\epsscale{1.0}
\label{m33_thumb2}
\end{figure}

\begin{figure}
\figurenum{4}
\epsscale{0.7}
\plotone{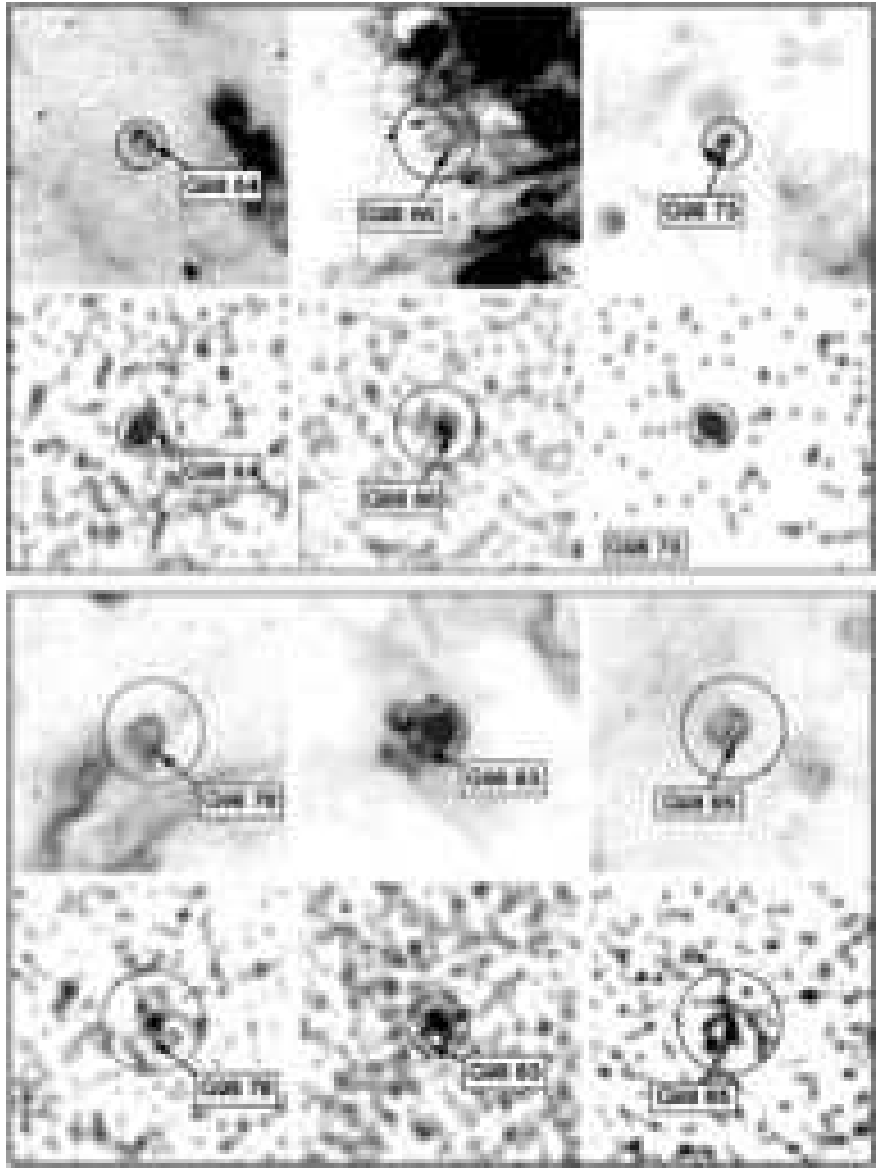}
\caption{Same as Figure \ref{m33_thumb1}.  All thumbnails measure
1\farcm5 on the side.  A full resolution
figure can be found at http://fuse.pha.jhu.edu/\~{}parviz/papers/m33/. }
\label{m33_thumb3}
\end{figure}

\begin{figure}
\epsscale{0.7}
\figurenum{5}
\plotone{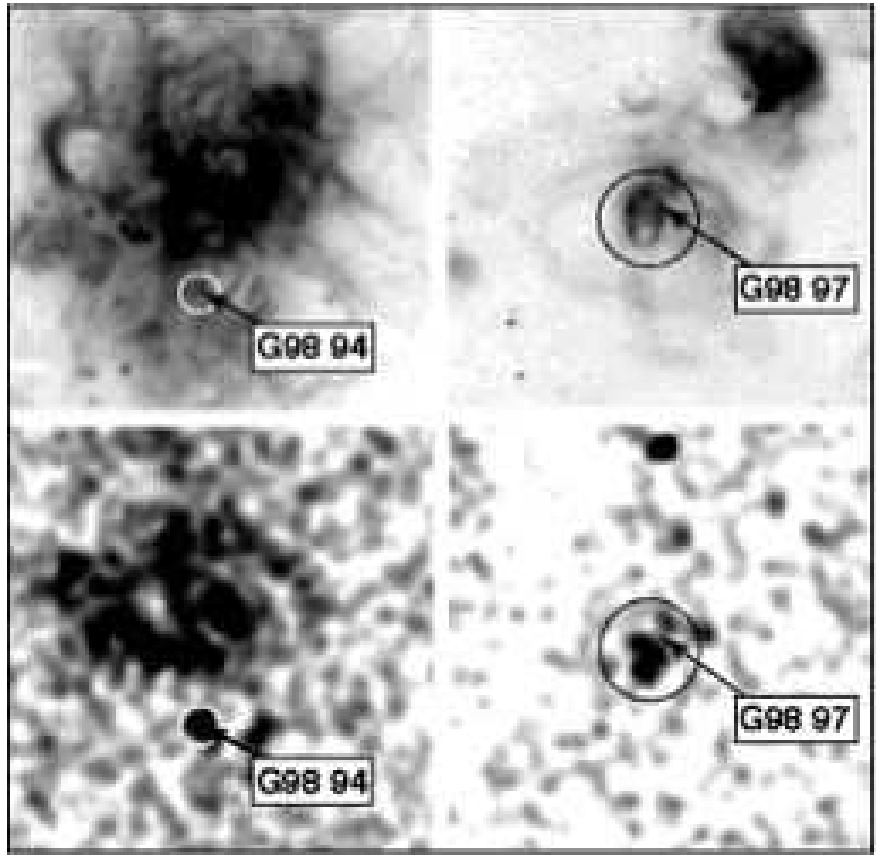}
\caption{Same as Figure \ref{m33_thumb1}.  SNR 94 is located inside the starburst
H~II region NGC 604.  All thumbnails measure
1\farcm5 on the side.  A full resolution
figure can be found at http://fuse.pha.jhu.edu/\~{}parviz/papers/m33/. }
\label{m33_thumb4}
\end{figure}

\begin{figure}
\figurenum{6}
\plotone{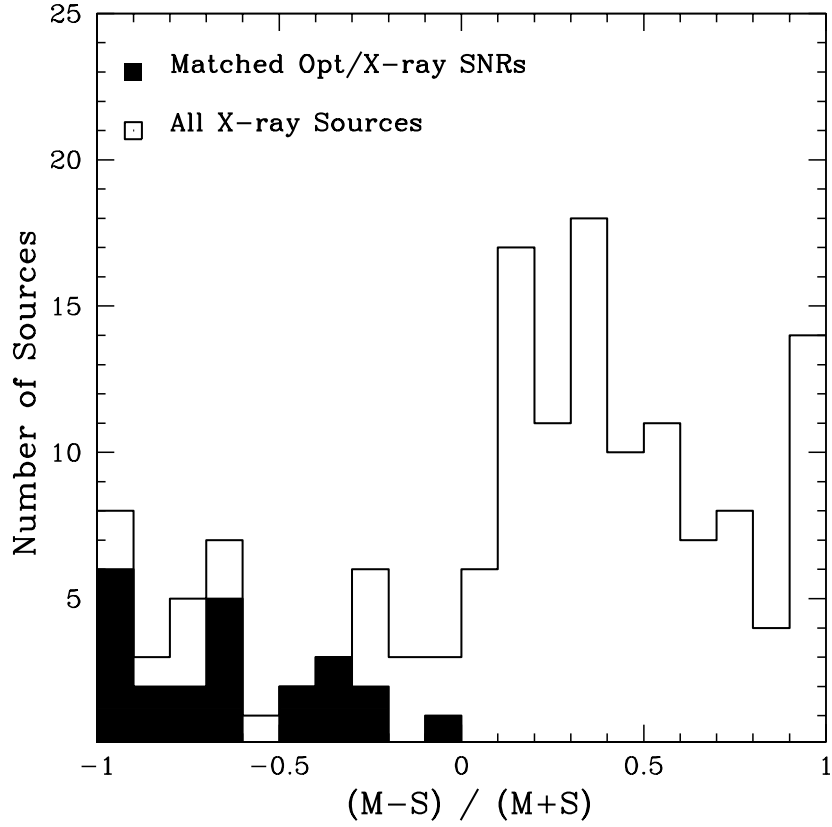}
\caption{
Hardness ratios for the 166 detected sources ($>$ 3$\sigma$) in M33, where S and
M are defined as emission in the 0.35$-$1.1 keV and 1.1$-$2.6 keV range, respectively.
The hardness ratios of the 23 sources matching optically identified SNRs (22 objects from
GKL98 and 1 soft source matching a newly identified optical SNR; see Section 8) are also
shown for comparison.  }
\label{m33_hardratios}
\end{figure}

As shown in Figure \ref{m33_halpha}, there appears to be a cluster of X-ray SNRs detected in the southern spiral arm of
M33, just south of the nucleus.  This may be a selection effect due to the coverage of the spiral
arm and nucleus by the ACIS S3 chip in the ACIS-S imaging dataset (ObsID 786).  Although the same region
is covered in an ACIS-I imaging observation (ObsID 1730) with nearly the same integration time, the factor of 2
higher sensitivity of the S3 chip results in a larger number of SNR detections (19 optically known SNRs
covered by the S3 chip, 7 detected).  

Guided by the spectral softness of the SNR candidates, we calculated luminosities for the sources under the assumption
that their spectra are dominated by thermal emission attenuated by M33 and Galactic absorption.  Aside from 
remnants 28, 29, 31, 35 and 55, the remaining objects exhibit $<$\,100 counts in their broadband (0.35$-$8.0 keV)
spectra (Table~2).   In these fainter sources it is not possible to detect and/or resolve any of the emission line 
structure one expects from a thermal plasma.  Therefore, the only
quantities we can meaningfully constrain in these sources are those that influence the overall shape of the 
thermal spectrum such as the plasma temperature ($kT$), the local absorbing column (N$_{H}$) and spectral normalization.

\begin{figure}
\figurenum{7}
\includegraphics[height=5.0in,angle=270]{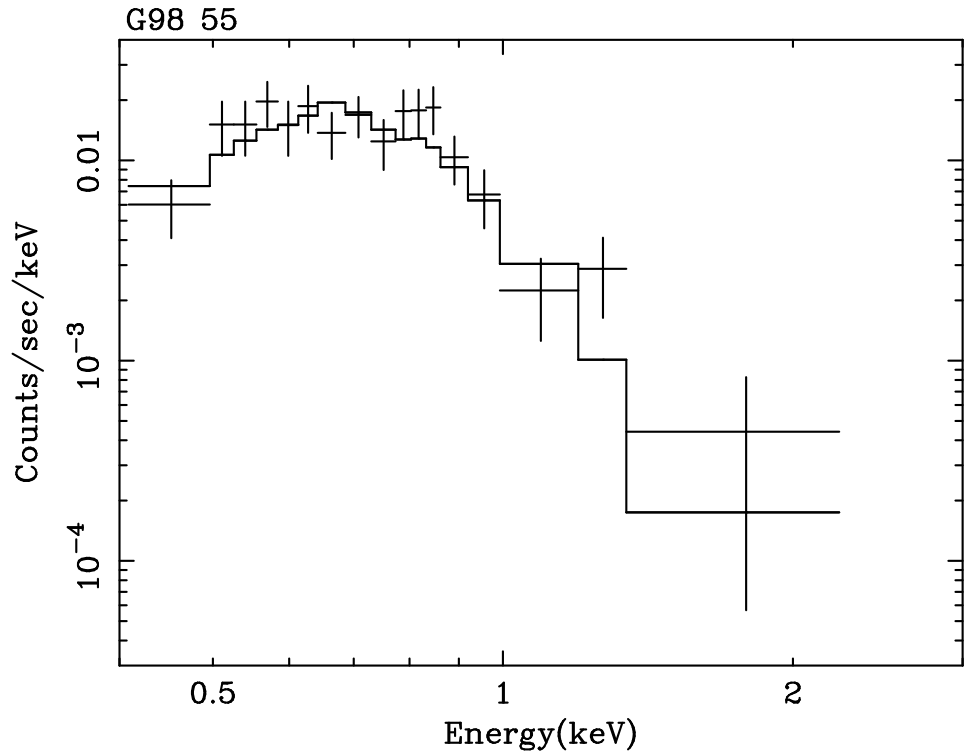}
\caption{X-ray spectrum of SNR 55 from GKL98, the brightest X-ray SNR detected in the \chan\,
observations of M33.  The best fit Raymond-Smith model is marked by the solid line drawn
through the data points.
}
\label{snr55_fit}
\end{figure}

Given the above constraints, we adopted the following approach for estimating the luminosities of the SNR candidates.
First, we generated ancillary response files (arfs) and redistribution matrix files (rmfs) for each object.
This accounted for the spatial and energy dependences of the High Resolution Mirror Assembly (HRMA) and
ACIS responses at the position of each source.
Then we fit the \chan\, spectrum of the brightest SNR candidate in our
sample (number 55, 480 integrated counts) with a Raymond-Smith model in XSPEC 11.3.1.  Quantities held fixed
during the fit were the Galactic absorbing column (N$_{H}$(Gal)\,=\,5.4$\times$10$^{20}$ cm$^{-2}$, Stark 
\etal\, 1992) and the abundances of the emitting plasma.  We set the latter quantity to 0.4 solar, the average
value obtained by Blair \& Kirshner (1985) from optical spectroscopy of 12 SNRs in M33.
The local M33 absorbing column N$_{H}$(M33), plasma temperature $kT$
and spectrum normalization were left free.  After fitting the spectrum of remnant SNR 55 (Figure \ref{snr55_fit}),
we fit the spectra of the remaining sources 
by fixing N$_{H}$(M33) to the best value obtained for SNR 55 (N$_{H}$(M33)\,=\,7$\times$10$^{20}$ cm$^{-2}$) 
and using the $kT$ and normalization for SNR 55 as starting values for the fits.
The M33 column density obtained by the spectral fit of SNR 55 lies comfortably in the range of values
measured in H~I observations of that galaxy (5$\times$10$^{20}$\,$\leq$\,N$_{H}$(M33)\,$\leq$\,5$\times$10$^{21}$;
Newton 1980), giving confidence that the columns used in our fits are reasonable (we assume solar metallicity
for the M33 absorbing column; a metallicity of 0.4 solar would raise N$_{H}$(M33) to 1.3$\times$10$^{21}$ cm$^{-2}$).
Finally, we converted the resulting intrinsic fluxes to luminosities assuming a distance of 795
kpc for M33.  The best fit temperatures and calculated luminosities are shown in Table~3.  The faintest optically matched SNR
in the \chan\, field of view is associated with SNR 37 from the GKL98 catalogue and
exhibits a 0.35$-$8.0 keV luminosity of 7$\times$10$^{34}$ \ergss ($S_{X}$\,$\approx$\,9$\times$10$^{-16}$
ergs cm$^{-2}$ s$^{-1}$).  This is comparable to the depth achieved for older surveys of SNRs in
much nearer galaxies such as the Large Magellanic Cloud (LMC) (Long, Helfand \& Grabelsky 1981), and demonstrates that
close to the optical axis ($\lesssim$\,4\arcmin) a pointed ACIS-S imaging observation of M33 is capable of detecting
some of the faintest M33 remnants in a 50 ks integration.

\subsection{Comparison with the \xmm\, Source Catalogue}

An X-ray survey of M33 has recently been completed with \xmm\, (Pietsch \etal\, 2003, 2004).  
The survey covered the D$_{25}$ isophote of the galaxy (nearly twice the area observed by
\chan) and to a relatively uniform depth L(0.2-4.5 keV)\,=\,10$^{35}$ ergs s$^{-1}$). In their
analysis of the \xmm\, data Pietsch \etal\, (2004) identified 408 sources in M33, with 28 of the sources matching
known optical SNRs from GKL98 (Table~3 of Pietsch \etal\, 2004) and exhibiting soft spectra characteristic of shock heated gas. 
Pietsch \etal\, (2004) found counterparts to 13 SNRs from the GKL98 catalogue not detected in our \chan\, analysis:
SNRs 11, 12+13 (blended), 15, 18, 20, 21, 25, 36, 57, 59, 86+87 (blended).  These remnants are
not detected by \chan\, because they are either intrinsically too faint, located too far off the
optical axis (rendered undetectable by such effects as smearing
of the PSF and/or high detector background) or are located outside the \chan\, field of view.

Although the portion of M33 covered by the \chan\, observations lies completely within
the larger field covered by \xmm, the \chan\, observations were still able to detect
X-ray counterparts to 6 optical SNRs not revealed by the \xmm\, survey (Table~1).
This may be due to the location of some of these SNRs within
regions of extended diffuse emission.  If the source is faint enough, the
larger PSF of \xmm\, (10\arcsec) can cause SNR emission to merge with the surrounding diffuse background, resulting
in a non-detection.
An excellent example is the case of SNR 94 from the GKL98 catalogue (Figure~\ref{m33_thumb4}).
This remnant is embedded within the starburst H~II region NGC 604 (D'Odorico, Dopita \& Benvenuti 1980, GKL98),
and is surrounded by diffuse X-ray emission.
The source is well detected in the \chan\, data (S/N\,$\approx\,$4), but is
absent from the \xmm\, source catalogue.  This is a clear illustration of the advantages of \chan\, over \xmm\, in
locating X-ray sources in confused regions.
Most of the optical SNRs with X-ray counterparts are also known radio sources (Gordon \etal\, 1999).  As shown in Table~1, 
most of these objects exhibit negative radio spectral indices.

\section{Non-Radiative SNR Candidates}

Apart from the 22 SNRs detected in our sample, there are 84 soft sources in the \chan\, data which do not
match any optically identified SNRs, known foreground stars, X-ray binaries (XRBs) or supersoft sources.  
While some of these objects may be
background AGN, we can also expect that some will be young, non-radiative SNRs expanding in low
density environments.  The spectra of these soft sources contain too few counts ($\lesssim$\,100) for
meaningful X-ray fits, making the task of uniquely identifying them from the \chan\, data alone impossible.

Despite the above limitations, we can at least identify potential non-radiative SNR candidates in M33
by searching for radio counterparts to the soft X-ray sources.  The radio continuum emission from 
active stars, supersoft sources and XRBs is intrinsically faint (see Ogley \etal\, (2002) for radio
observations of Galactic supersoft sources and Fender, Southwell \& Tzioumis
(1998) for a radio survey of transient sources in the Magellanic Clouds), while synchrotron
radiation from particle acceleration can produce radio fluxes of up to a few mJy at the distance of
M33 (Gordon \etal\, 1999).   Note that the method of selecting SNR candidates is limited by the differing
sensitivities of the X-ray and radio data, since an unknown number of non-radiative SNRs may be intrinsically 
faint in the radio and remain undetected in the Gordon \etal\, (1999) observations.  

We also performed a cross-correlation between our broad-band source lists and Gordon \etal\,
(1999) radio sources not matching any known optical SNRs.  We found no matching radio sources 
to within 15\arcsec\, of any broad-band object.  However, a similar comparison with our soft
X-ray source lists revealed one matching source in ObsIDs 786 and 1730, with the former observation
showing the stronger detection (S/N\,=\,10).  The broad band, background-subtracted count rate of the source
from ObsID 786 is 2.7$\times$10$^{-3}$ cts s$^{-1}$.
The source is located at $\alpha_{J2000}$\,=\,01$^{\rm h}$33$^{\rm m}$50.5$^{\rm s}$,
$\delta_{J2000}\,=\,$30$^{\circ}$38\arcmin\,21\farcs5 and lies within 0.7\arcsec\, of radio source 100 from the Gordon \etal\, 
(1999) catalogue.  Following the CXC naming convention, we call this source CXOM33 J013350.5+
3038215.  This source lies approximately 75\arcsec\, south of the X-ray nuclear source (X-8), at
the base of the inner spiral arm.  There is little nebular emission visible within 20\arcsec\, of the X-ray
source.  The 6-20 cm spectral index of CXOM33 J013350.5+3038215 is consistent with synchrotron emission from shocks 
($\alpha_{6-20}$\,=\,
$-$0.6$\pm$0.1), but appears to exhibit a somewhat harder spectrum than remnants with optical counterparts
(from our \chan\, analysis HR1\,=\,$-$0.06$\pm$0.3, HR2\,=\,$-$0.33$\pm$0.35).  With only 126 counts in the spectrum, we are unable
to establish whether the emission originates in a combination thermal/non-thermal plasma, or whether the
hardness ratio is attributable to line emission from metal-rich ejecta.  The radio spectral index is also consistent
with emission from a background radio galaxy, though no obvious optical emission is seen at the position of the X-ray
source in the R band image of M33.  Despite the strong
detection of this object in the \chan\, observation (S/N\,$\sim$\,10), it is not detected in the \xmm\, data due to
its proximity to the nearby bright source M33 X-8.  We examined
the light curve of the source using data from ObsIDs 786 and 1730 and found that it exhibited no obvious variability to
within the errors. CXOM33 J013350.5+3038215 does not show clear evidence of an extended morphology.  It is 
clearly a candidate for a non-radiative SNR and a prime example of an object which would benefit from deeper
followup X-ray observations of M33.  

\section{Search for Optical SNR Emission from Soft X-ray Sources}

\subsection{Optical Emission from \chan\, Sources}

Given the richness of the \chan\, datasets, it is tempting to use the  
X-ray source lists as a guide to finding SNRs in optical narrowband images of M33.  Our approach
was to overlay the positions of soft band X-ray sources lacking counterparts from the GKL98
catalogue onto the NOAO Mosaic H$\alpha$, [S~II] and [O~III] images of M33.
We visually examined each field
where X-ray sources were located, searching for telltale signs of SNR emission such as filaments
and shell-like structures.  We also generated [S~II]/H$\alpha$ ratio images from the Mosaic data to
search for enhancements in the ratio expected from radiative shocks.  

\begin{figure}
\figurenum{8}
\epsscale{0.7}
\plotone{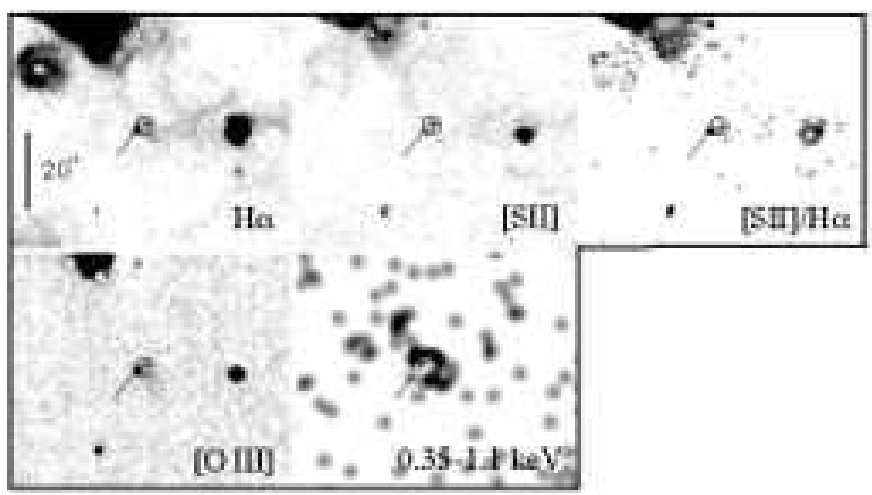}
\caption{Closeup view of the optical knot (indicated by the pointer) associated with \xmm\, source 270.  The X-ray source
was proposed as a SNR candidate by Pietsch \etal\, (2004) on the basis of its hardness ratio.  
The [S~II]/H$\alpha$ ratio of the knot 0.7 to 0.8, confirming that it is shock excited and that \xmm\,
source 270 is a SNR.  The circle
marks the \xmm\, position of the X-ray source, while the cross point marks the position measured from the
soft band (0.35$-$1.1 keV) \chan\, image.  The size of each mark indicates the approximate positional uncertainty
of the \xmm\, and \chan\, positions.  The X-ray image has been smoothed with a 3-pixel Gaussian.  A full resolution
figure can be found at http://fuse.pha.jhu.edu/\~{}parviz/papers/m33/.}
\label{snr100}
\end{figure}

Our comparison of the optical and X-ray images of M33 revealed one new optical counterpart:
an emission knot exhibiting an elevated [S~II]/H$\alpha$ ratio (0.7$-$0.8; fully consistent with radiative shock excitation) and matching a 
soft source located 6\farcm8 north of the starburst H~II region NGC 604 (Figure~\ref{snr100}).  The X-ray source
is detected in ObsID 2023, and exhibits hardness ratios consistent with thermal emission:
HR1\,=\,$-$0.36$\pm$0.5, HR2\,=\,$-$0.13$\pm$0.74.  It matches the position of \xmm\, source 270, one
of 16 X-ray sources detected by \xmm\, and classified as a SNR by Pietsch \etal\, (2004) on the basis
of hardness ratios.  This source is located at $\alpha_{J2000}$\,=\,01$^{\rm h}$34$^{\rm m}$23.3$^{\rm s}$,
$\delta_{J2000}\,=\,$30$^{\circ}$54\arcmin\,24\farcs0 in the soft band image of M33, agreeing with the \xmm\, position to well within
the positional uncertainty of both observations ($\sim$1\farcs5$-$2\farcs0).  This source, which we designate
CXOM33 J013441.0+3043280, exhibits only 30 counts in ObsID 2023
and lies well off the imaging axis of \chan, so we are unable to determine whether it is extended.
In addition, there is no radio counterpart to CXOM33 J013441.0+3043280 from the catalogue of Gordon \etal\, (1999).
Assuming a Raymond-Smith plasma with model fitting parameters from Table~3, we obtain a temperature of
0.3 keV and a 0.35$-$8 keV luminosity \lx\,=\,2.3$\times$10$^{35}$ \ergss.
Although the portion of M33 containing CXOM33 J013441.0+3043280 was covered by GKL98, identification
of the optical counterpart was not included in that survey.

\subsection{Optical Emission from \xmm\, Sources}
  
Of the 16 \xmm\, sources categorized as SNRs by Pietsch \etal\, (2004), only 3 were detected
at the 3$\sigma$ level in the soft band \chan\, images (one of these objects is source 270 described
above).  The remaining 13 \xmm\, SNR candidates
either fell below the detection threshold of \chan\, or lay outside the field
of view of the \chan\, images.  However, we can at least search for optical counterparts to these
candidates in the KPNO Mosaic images.

\begin{figure}
\figurenum{9}
\epsscale{0.7}
\plotone{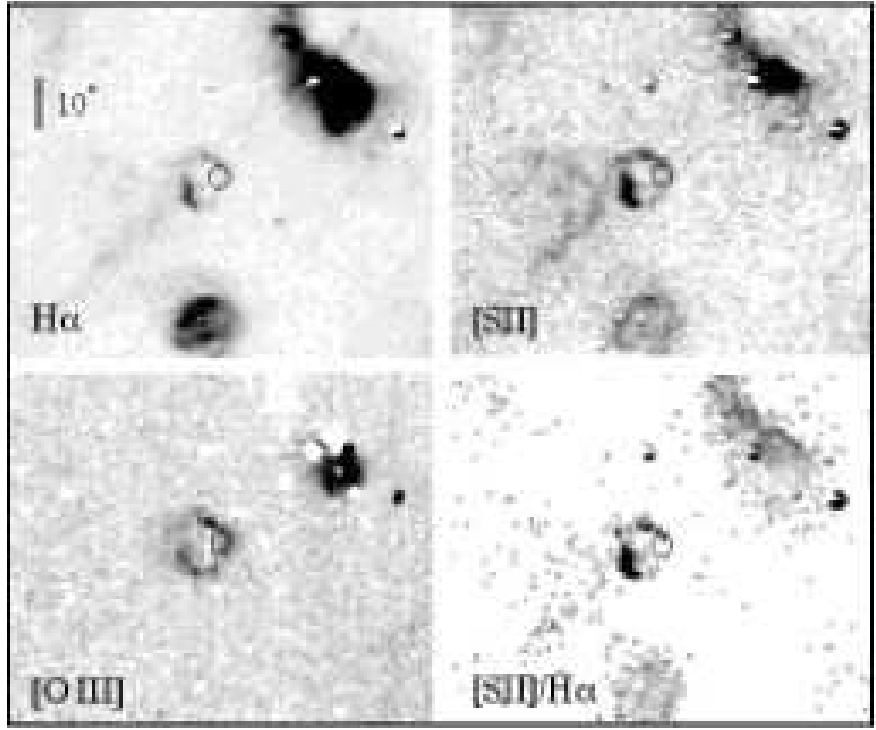}
\caption{Closeup view of the optical shell associated with \xmm\, source 68.  The X-ray source
was proposed as a SNR candidate by Pietsch \etal\, (2004) on the basis of its hardness ratio.  
The source was not detected by \chan\, due to its faintness and far off-axis position.
The [S~II]/H$\alpha$ ratio along the shell varies from 0.6 to 0.8, confirming that the shell
is shock-excited and that \xmm\, source 68 is a SNR.  A full resolution
figure can be found at http://fuse.pha.jhu.edu/\~{}parviz/papers/m33/. }
\label{xmm68}
\end{figure}

Overlaying the positions of the 13 \xmm\, SNRs onto the continuum-subtracted H$\alpha$, [S~II] and
[O~III] images, we found a good match between \xmm\, source 68 and an optical shell 9\arcsec\,
across (35 pc), located near the end of the southern spiral arm at $\alpha_{J2000}$\,=\,
01$^{\rm h}$32$^{\rm m}$46.5$^{\rm s}$,
$\delta_{J2000}\,=\,$30$^{\circ}$34\arcmin\,39\farcs0.  The eastern side of the shell
is particularly bright (Figure \ref{xmm68}) and exhibits a [S~II]/H$\alpha$ varying from 0.6 to 0.8, confirming
that the emission is shock excited.   The shell is also detected in [O~III], where its emission is more
evenly distributed along the rim.  The \xmm\, hardness ratios for source 68 are
$(M\,-\,S)/(M\,+\,S)\,=\,$0.14$\pm$0.16 and $(H\,-\,M)/(H\,+\,M)\,=\,-$0.79$\pm$0.18, where
$S$ is (0.2$-$0.5) keV, $M$ is (0.5$-$1.0) keV and $H$ is (1.0$-$2.0) keV.

Like \xmm\, source 270, source 68 escaped identification in the optical survey of GKL98
and exhibits no radio counterpart in the Gordon \etal\, (1999) catalogue.
This object lies in the field of \chan\, ObsID
1730, approximately 14\arcmin\, off axis on the S3 chip.  The faintness of this source
(F$_{X}$(0.1$-$4.5 keV)\,$\sim\,$10$^{-15}$ ergs cm$^{-2}$ s$^{-1}$ in the \xmm\, data, Pietsch \etal\, (2004))
along with the strongly broadened \chan\, PSF at its off-axis position are likely responsible for the lack of
detection of this source in the \chan\, data.

Since the H$\alpha$ filter used in the
M33 imagery also transmits [N~II] line emission, we have likely underestimated the [S~II]/H$\alpha$ ratios
of CXOM33 J013441.0+3043280 and \xmm\, source 68.  Optical spectroscopy will be required to obtain a more accurate measurement and
to better characterize the physical properties of this SNR.   The discovery of optical emission from CXOM33 J013441.0+3043280
and \xmm\, source 68 brings the total number of confirmed SNRs in M33 to 100.
The total number of unique remnants with optical counterparts in both the \chan\, and \xmm\, observations is 37,
nearly 1/3 of the total identified in the optical.

\section{Properties of Optical SNRs Detected with \chan\, and \xmm}

\begin{figure}
\figurenum{10}
\plotone{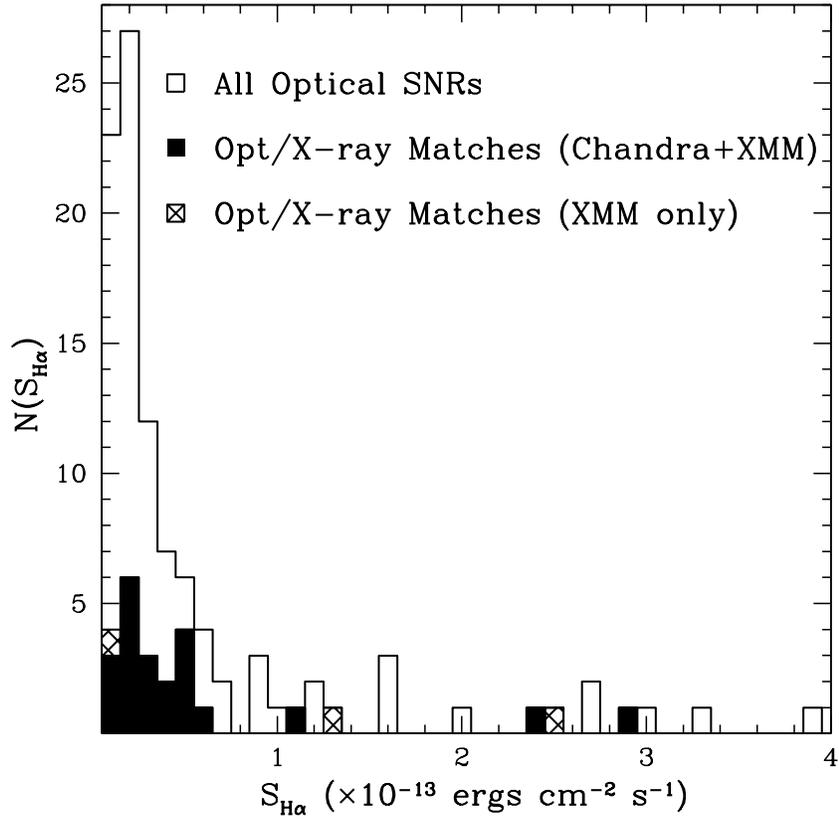}
\caption{The number of known optical SNRs per H$\alpha$ flux interval (from the
catalogue of GKL98).  The same relation is also plotted for the subset
of optical SNRs exhibiting X-ray emission as identified by \chan\, and \xmm. }
\label{m33_hafluxes}
\end{figure}

Although there are insufficient counts in the spectra of the SNR candidates to obtain detailed
information on these sources, we can at least explore correlations between the X-ray and optical
observations and identify systematic trends which may reveal properties of the global SNR
population in M33.
As a first test, we searched for selection effects between the brightnesses of optically identified SNRs
and the number of such remnants detected in the X-rays.   Assuming that each SNR is of 
uniform optical surface brightness (an admittedly rough assumption), we calculated H$\alpha$ fluxes for each object
using the H$\alpha$ surface brightnesses and physical sizes of SNRs listed in the GKL98 catalogue.
In Figure \ref{m33_hafluxes}\, we present a histogram showing the number of optical SNRs $N(S_{H\alpha})$ per
H$\alpha$ flux interval  $S_{H\alpha}$ (98 SNRs from GKL98), with the 22 GKL98 SNRs exhibiting
X-ray emission marked separately for comparison.  For completeness we have also marked
the H$\alpha$ fluxes of optical SNRs detected by \xmm.

It is clear from Figure \ref{m33_hafluxes}\, that the brightest optical SNRs in M33 are not preferentially
detected in the \chan\, observations.  
Rather, the greatest number of X-ray detected SNRs are found where the greatest
number of optical SNRs are found: at the optically faint end of the $N(S_{H\alpha})$ vs. $S_{H\alpha}$
relation.   This is markedly different from what is observed in SNR studies of other spirals
such as M83 (Blair \& Long 2004).  Although M83 was observed with \chan\, for a similar exposure time to 
M33 (50 ks), the greater distance of that galaxy resulted in an X-ray survey depth \lx(0.3$-$8.0 keV)\,$\gtrsim$\,
10$^{36}$ \ergss (Soria \& Wu 2003), nearly an order of magnitude shallower than the M33 observations.  Blair \&
Long (2004) noted that this sensitivity-induced selection effect
likely resulted in the detection of only the brightest X-ray SNRs in M83.  Interestingly, Blair \& Long (2004) 
stated that the systematic detection of the brightest optical SNRs as X-ray sources in M83 may also be
due to the expansion of these remnants into denser than average regions.  In that case, the lack of
such a strong trend in M33 (say, for remnants with $S_{H\alpha}\,\geq$\,5$\times$10$^{-14}$ ergs cm$^{-2}$ s$^{-1}$,
where only 10 out of 32 optical remnants are detected in the X-rays by \chan\, and \xmm) may be partially caused by
expansion of the SNRs into regions of lower than average density.  

\begin{figure}
\figurenum{11}
\plotone{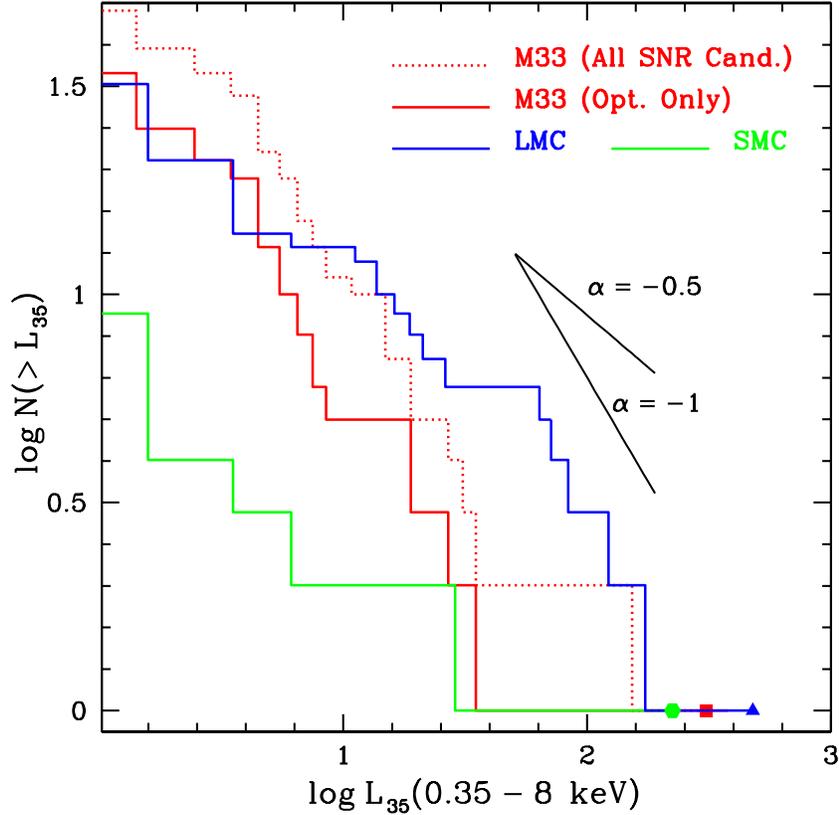}
\caption{The cumulative luminosity distribution of X-ray sources (combined \chan\, and \xmm\,
sample) matching optically identified SNRs in M33.   Two luminosity distributions are shown
for M33: the solid red histogram shows the 37 SNRs exhibiting optical counterparts (including
all optical SNRs detected by both \xmm\, and \chan), while the dotted red histogram shows
both the optically confirmed SNRs and X-ray SNR candidates identified by \xmm\, (Pietsch \etal\, 2004)
(51 sources).  The luminosity distributions of 46 LMC SNRs observed by \rosat\, (Sasaki, Haberl \& Pietsch 2000) and
13 SMC SNRs observed by \xmm\, (van der Heyden, Bleeker \& Kaastra 2004) are also shown for comparison.  The
brightest SNRs in the SMC, M33 and LMC respectively are marked by the pentagon, square and triangle (corresponding
to E0102$-$72.3, GKL98 21 and N132D).  Lines of slope $\alpha$\,=\,$-$0.5 and $-$1 are shown as a reference.  Only
SNRs brighter than 10$^{35}$ \ergss\, are included in this plot to facilitate visual
comparison between the high luminosity ends of the three distributions. }
\label{m33_lum}
\end{figure}

Another relationship we can measure from the \chan\, data is the cumulative luminosity distribution, $N(>L)$,
of the M33 SNRs in the X-rays. Comparing this distribution with that of other
extragalactic SNR populations observed
in the X-rays such as the LMC and SMC provides a comparative 'snapshot' of each SNR sample.  Since
the derived luminosity depends upon the assumed spectral model, we have attempted to reduce the systematic
differences in the tabulated luminosities of extragalactic SNR surveys
by applying the same spectral model when converting from instrument-specific
count rates to fluxes.  Utilizing both \chan\,
count rates from our M33 SNRs (Table~2) and count rates for the optically matched SNRs detected
exclusively in the \xmm\, observations of M33 (Pietsch \etal\, 2004), along with count rates for the LMC SNRs
(46 sources from the \rosat\, catalogue of
Sasaki, Haberl \& Pietsch (2000)) and count rates for the SMC SNRs (13 sources from the \xmm\, survey of van der Heyden,
Bleeker \& Kaastra 2004), we obtained luminosities in the range 0.35$-$8.0 keV for each SNR distribution.
We utilized the PIMMS online tool of CXC for the calculations and assumed a Raymond-Smith
model, $kT$\,=\,1 keV and 0.2 solar abundances.  We fixed the M33 column to the
best fit value for SNR 55 (N$_{H}$(M33)\,=\,7$\times$10$^{20}$ cm$^{-2}$, as found in Section 5.2) 
and used N$_{H}$(LMC)\,=N$_{H}$(SMC)\,=\,2$\times$10$^{20}$ cm$^{-2}$,
(Heiles \& Cleary 1979).  The Galactic column used for all
three galaxies was N$_{H}$(Gal)\,=\,5$\times$10$^{20}$ cm$^{-2}$.  The assumed distances to M33, the LMC and
SMC were 795, 50 and 60 kpc, respectively.
In Figure \ref{m33_lum}\, we present a histogram of log$\,N(>L)$ vs. log$\,L(0.35-8.0)$ for all M33 SNRs
exhibiting optical counterparts, along with histograms for SNRs in the LMC and SMC (note that in
calculating SNR luminosities for all three galaxies under uniform assumptions we arrive at
different M33 luminosities than the (more accurate) values reported in Table~3).

A comparison of the luminosity relations of the three SNR samples in Figure \ref{m33_lum}\, 
reveals several interesting features.  First, the 0.35$-$8.0 keV luminosity of the brightest GKL98
remnant detected by \xmm, SNR 21, is quite high: $\sim$3$\times$10$^{37}$
\ergss\, (in good agreement with the 0.1$-$2.4 keV value measured from \rosat\, PSPC observations by
Long \etal\, (1996)).   By comparison, the two young ejecta-dominated remnants E0102$-$72.3 in the SMC
and N132D in the LMC exhibit luminosities $\sim$2$\times$10$^{37}$ \ergss\, and $\sim$4$\times$10$^{37}$ \ergss,
respectively.  SNR 21 exhibits a large optical diameter (28 pc, GKL98) and features an optical spectrum
dominated by emission from shocked interstellar gas (Smith \etal\, 1993). It is undoubtedly
older and more evolved than the two brightest Magellanic Cloud remnants.  

Another noticeable feature of Figure \ref{m33_lum}\, is the clear luminosity separation
between the M33, SMC and LMC distributions.  There appear to be fewer SNRs at the high
luminosity end of the M33 distribution ($>$\,10$^{36}$ \ergss) than in the LMC:
while there are a similar number of confirmed X-ray SNRs in M33 and the LMC with X-ray luminosities
exceeding 10$^{35}$ \ergss, nearly 40\% of the LMC SNRs are brighter than 10$^{36}$ \ergss, while only 13\% of
the M33 sample exceed this luminosity.  The opposite trend is seen between M33 and the SMC, although
the relatively smaller sample size of the SMC SNRs makes the comparison of its luminosity function to
those of the LMC and M33 more uncertain.   The offset in brightness between the M33 and Magellanic Cloud SNRs was
also noted by Haberl \& Pietsch (2001) in their \rosat\, analysis of M33,
although the larger sample of remnants in our comparison (35 remnants rather than 13) has filled
in more of the low luminosity population, reducing the contrast between the M33 and LMC distributions.
However, the question remains: are the luminosity offsets between the three distributions real?  The remnant 
to remnant luminosity differences are small enough ($\sim$2$-$5) to be
accounted for by uncertainties in temperature and column density: varying the temperatures from 0.5 keV
to 3 keV causes a factor of 2 variation in the calculated luminosities, while varying the column densities
over the full range allowed for each galaxy produces a factor of 3 variation in calculated luminosities.
Allowing the abundances to range from 0.2 to 0.5 solar yields a 10\% variation in luminosity. 
However, it is unlikely that variations in these three parameters
would combine in just the right proportion to systematically lower the luminosities of the brightest M33 SNRs
to values below those of the brightest LMC SNRs.  Likewise, it is also unlikely that the variations in
emission parameters would systematically raise the M33 luminosity distribution above
that of the SMC. 

Aside from intrinsic physical causes, another factor influencing the luminosity distributions is the 
relatively lower completeness of
the M33 SNR sample compared with remnants in the SMC and LMC.  Thus far we have discussed
only the M33 SNRs with known optical counterparts, while the SMC and LMC samples include pure non-radiative
SNRs.  Clearly some of the optically invisible, soft X-ray sources detected in the \chan\, observations of M33
may be non-radiative SNRs.  If these remnants are merely unidentified, they could in principle be bright enough to fill
in the high end of the M33 luminosity distribution.  To test the potential influence of non-radiative SNRs
on the M33 luminosity function, we note that of the 16 soft sources classified as SNRs by \xmm\, purely
from hardness ratios, 2 objects (sources 68 and 270) have been shown in this paper to exhibit optical emission.  The remaining
14 objects are candidates for non-radiative SNRs, so in Figure \ref{m33_lum}\, we have included a second
histogram for M33 showing the luminosity distribution with these sources added in (total of 51 SNRs and SNR
candidates).  Here we have computed the luminosities of the 14 X-ray SNR candidates using the same
Raymond-Smith model parameters as the optically confirmed SNRs.   

Comparing the luminosity functions
of the optically confirmed SNRs (solid red histogram in Figure \ref{m33_lum}) with those of the combined
sample (dotted red histogram), it is clear that aside from shifting the M33 distribution upward in log$\,N(>L)$,
the added X-ray SNR candidates do fill in some of the M33 distribution at high luminosities as expected,
slightly reducing the slope of the distribution above 10$^{36}$ \ergss\, (now the fraction of M33 SNRs
brighter than 10$^{36}$ \ergss\, has increased from 13\% to 22\%). However, there is still a conspicuous
gap between the M33 and LMC distributions.
We note that some of the \xmm\, SNR candidates may be supersoft sources with spectra hardened enough by          
local absorption to mimic SNR emission.  However, given the small number of supersoft sources
detected in the \xmm\, survey (5 objects out of a total of 408; Pietsch \etal\, (2004)) it seems unlikely
that the sample of \xmm\, SNR candidates is significantly contaminated with heavily absorbed supersoft sources.
Nevertheless, the danger remains that in principle some of the \xmm\, candidates may be misclassified objects.

There is another selection effect introduced by the \chan\, observations which further reduces the completeness
of the M33 luminosity distribution.
In our analysis we have not taken into account the effects of the decreasing sensitivity of \chan\,
with off-axis angle.  As noted earlier, the broadening of the
\chan\, PSF away from the aimpoint raises the detection threshold for X-ray sources, resulting
in a progressively incomplete luminosity distribution at farther off-axis angles (Kim
\& Fabbiano 2003).  While this clearly reduces the number of SNRs at the faint end of the M33 distribution, it
cannot account for the clear separation between the M33 and the LMC SNRs at high ($\gtrsim\,$10$^{36}$
\ergss) luminosities.  The overall conclusion of our analysis is that even allowing for uncertainties
in spectral parameters and the incompleteness of the M33 supernova remnant sample, a real difference likely exists between
the luminosity distributions of M33, the SMC, and the LMC.   

The separations between the three distributions may reflect real physical differences.  Haberl \& Pietsch (2001) 
attributed the high luminosities of the LMC SNRs to the particularly high metallicity of the interstellar medium in the LMC.
By that line of reasoning, the progressively fainter luminosity distributions of M33 and the SMC may
be caused by progressively lower metallicities in those galaxies.  Previous measurements of interstellar
abundances in the three galaxies are consistent with this interpretation: the 0.4 solar abundances estimated
from spectra of the optical SNRs in M33 (Blair \& Kirshner 1985) lie between the 0.2 and 0.5 solar abundances
measured by Russell \& Dopita (1992) for the SMC and LMC, respectively.  However, as mentioned earlier, 
abundance changes in the context of our assumed model are not sufficient to explain the luminosity
offsets seen in Figure \ref{m33_lum}.  Other factors
such as differences in average interstellar medium density and explosion energy may also play a role
in separating the three luminosity distributions, but a full exploration of these effects is beyond
the scope of this paper.

\section{Summary}

We have performed the first systematic search for \chan\, X-ray counterparts to optically identified SNRs
in M33.  Aside from matching X-ray sources with known optical SNRs, we have also attempted to use
soft sources detected by \chan\, to find optical counterparts missed by earlier narrowband imagery of
M33.  Our search was performed using \chan\, archival images of M33 and narrowband H$\alpha$ and [S~II]
KPNO Mosaic images from the NOAO archive.  Our results are as follows:

1. We have found 22 X-ray counterparts to known optical SNRs from the GKL98 catalogue.
These X-ray sources exhibit soft spectra characteristic of thermal emission from shock-excited gas.
Comparing the \chan\, images of the SNR candidates with simulated images of point sources, we
have concluded that at least four of the SNR candidates exhibit an extended morphology.

2. One soft X-ray source in the \chan\, data exhibits no optical emission, but matches a
radio source with a steep spectral index suggestive of particle acceleration in shocks.  We propose that
this source may be a young non-radiative SNR, similar to SN 1006 in the Milky Way.

3. Compared to the optical/X-ray survey of the more distant galaxy M83 (5 Mpc), we find no evidence that the 
brightest optical SNRs are preferentially detected in the X-rays.  This
is likely due to a combination of higher completeness of the M33 survey caused by the close proximity of
that galaxy and a lower average interstellar medium density in M33.  

4. We have searched the KPNO Mosaic images for optical counterparts to both soft \chan\, sources
and objects identified by Pietsch \etal\, (2004) as
potential SNRs on the basis of \xmm\, hardness ratios.  In the \chan\, data we find
a positional coincidence between one soft source (matching object 270 from the \xmm\, catalogue of Pietsch \etal\, 2004) and
a knot of optical emission.  The knot exhibits a [S~II]/H$\alpha$ ratio of 0.7$-$0.8, strongly suggesting that
the emission from this object is produced in radiative shocks.  In addition, we find a positional coincidence between
one \xmm\, SNR candidate (object 68 from the \xmm\, catalogue) and a prominent optical shell approximately
9\arcsec\, across.  The [S~II]/H$\alpha$ ratio of this shell varies from 0.6 to 0.8, indicating that like source 270,
the emission from this object is produced in radiative shocks.  Followup optical spectroscopy will be
required to confirm the high [S~II]/H$\alpha$ ratios of the two newly confirmed SNRs.
However, we are fairly confident that the number of optically emitting SNRs in M33 is now 100.
The total number of soft X-ray sources matching optically known SNRs, including those found
in the \xmm\, survey of M33 (Pietsch \etal\, 2003, 2004), is 37 objects.  Nearly 1/3 of the
optical SNRs in M33 are now known to have X-ray counterparts.

5. We find that there are fewer confirmed, bright SNRs ($>$ 10$^{36}$ \ergss) in
M33 than in the LMC.  The opposite trend is seen when comparing the M33 and SMC distributions. 
This feature may partly reflect differences in interstellar abundances between the three
galaxies.  However, abundance differences do not fully account for the luminosity separation
between M33 and the LMC.  While adding X-ray SNR candidates (i.e., objects lacking obvious optical counterparts)
from the \xmm\, survey to the M33 luminosity distribution increases the total number of SNRs at all
luminosities and reduces the slope of the M33 distribution at the highest luminosities, the significant
separation between the M33 and LMC luminosity functions persists.  A full explanation of this
trend awaits future investigation.

The authors would like to thank H.-J. Grimm, J. P. Hughes and B. Sugerman for helpful discussions on the X-ray sources in M33,
and Gabriel Brammer for his assistance in the data analysis.  We would also like the thank the referee for assistance in
improving the presentation of this paper.  This work has been supported by NASA Grant AR3-4003B to JHU and
Grant AR3-4003X to STScI.  M. S., T. J. G. and P. P. acknowledge support under NASA contract NAS8-03060.

\begin{deluxetable}{cccccc}
\tabletypesize{\footnotesize}
\tablecaption{Identification of Optical/\chan\, SNR Matches in Radio and \xmm\, Catalogues}
\tablehead{
\colhead{ID\tablenotemark{a}} & \colhead{OBSID}  &  Radio ID\tablenotemark{b}  &  
\colhead{$\alpha_{6-20}$\tablenotemark{c}}  &  \colhead{XMM ID}\tablenotemark{d} & \colhead{Source Extended?}  }
\startdata
9  &   1730  	&  	11	&  	$-$0.5$\pm$0.3	&	93	&	inconclusive\tablenotemark{e}\\
27  &  1730  	&  	47	&  	$-$0.8$\pm$0.2	&	153	&	inconclusive\\
28  &   786  	&  	50	&  	$-$0.2$\pm$0.2	&	158	&	yes (marginal)\\
29  &	786	&	52	&	$-$0.9$\pm$0.5	&	161 	&	no \\
31  &	1730	&	57\tablenotemark{f}    	&	$-$0.8$\pm$0.1	&	164    &	no\\
35  &   1730	&	64	&	$-$0.7$\pm$0.1	&	179	&	yes (marginal) \\
37  &   786	&	none	&	\nodata		&	none	&	yes (marginal)\\
45  &   786	&	81	&	$-$0.6$\pm$0.2	&	none	&	yes\\
47  &   1730	&	90	&	$>$ $-$1.3	&	207	&	inconclusive\\
53  &   786	&	110	&	$>$ $-$0.8	&	213	&	inconclusive \\
54  &	1730	&	111	&	$-$0.9$\pm$0.1	&	214	&	inconclusive \\
55  &   786	&	112	&	$-$0.7$\pm$0.1	&	215	&	no\\ 
60  &	786	&	none	&	\nodata		&	none	&	no \\
62  &	786	&	125	&	0.0$\pm$0.3	&	225	&	no \\
64  &	786	&	130	&	$-$0.2$\pm$0.2	&	230	&	yes\\
66  &	786	&	none	&	\nodata		&	none	&	inconclusive\\
73  &	1730	&	148	&	$-$1.1$\pm$0.4	&	250	&	inconclusive\\
78  &	2023	&	none	&	\nodata		&	252	&	inconclusive\\
83  &	2023	&	157	&	$>$ $-$0.2	&	256	&	inconclusive\\ 
85  &	2023	&	160	&	$-$0.5$\pm$0.3	&	none	&	inconclusive \\
94  &	2023	&	none	&	\nodata		&	none	&	yes\\
97  &	2023	&	181	&	$-$0.1$\pm$0.2	&	314	&	yes\\
270\tablenotemark{g}  &	2023	&	none	&	\nodata		&	270	&	inconclusive\\
\enddata
\tablenotetext{a}{Optical IDS from the SNR catalogue of GKL98.}
\tablenotetext{b}{Radio IDs from the catalogue of Gordon \etal\, (1999).}
\tablenotetext{c}{Radio spectral index (6 cm - 20 cm) from catalogue of Gordon \etal\, (1999). }
\tablenotetext{d}{\xmm\, source catalogue of Pietsch \etal\, (2004). }
\tablenotetext{e}{These SNR candidates are faint and/or located too far off axis to allow
meaningful comparison between their angular extents and \chan\, PSF predictions.  See
text for more details.}
\tablenotetext{f}{The matching radio source from Gordon \etal\, (1999) lies 17\farcs2 away.}
\tablenotetext{g}{Source 270 from the \xmm\, catalogue matches a soft X-ray source in
the \chan\, observations.  See Section 7 for details. }
\end{deluxetable}

\begin{deluxetable}{cccccc}
\tabletypesize{\footnotesize}
\tablecaption{\chan\, X-ray Counterparts of Optical SNRs in M33}
\tablehead{
\colhead{ID\tablenotemark{a}} & \colhead{$\alpha_{2000}$\tablenotemark{b}}  & \colhead{$\delta_{2000}$} &  \colhead{Count Rate (10$^{-3}$ cts s$^{-1}$)\tablenotemark{c}}  &  
\colhead{(M$-$S)/(M+S)\tablenotemark{d}}  &  
\colhead{(H$-$M)/(H+M)}\tablenotemark{e} }
\startdata
9   &  01:32:57.1 &  30:39:24.8 &	0.60$\pm$0.19	&  $-$0.93$\pm$0.40  &  $-$1 \\
27  &  01:33:28.0 &  30:31:34.9 &  	0.38$\pm$0.16	& $-$0.96$\pm$0.43  &  $-$1 \\
28  &  01:33:29.0 & 30:42:17.1 &  5.24$\pm$0.34	&  $-$0.70$\pm$0.50      &  $-$0.97$\pm$0.24 \\
29  &  01:33:29.4 & 30:49:11.9 &  2.35$\pm$0.26	& $-$0.85$\pm$0.39      &  $-$0.50$\pm$0.99 \\
31  &  01:33:31.2 & 30:33:33.6 &  6.10$\pm$0.37 	& $-$0.35$\pm$0.66      &  $-$0.86$\pm$0.38 \\
35  &  01:33:35.9 & 30:36:28.0 &  2.10$\pm$0.22 	& $-$0.46$\pm$0.63      &  $-$1 \\
37  &  01:33:37.7 & 30:40:10.2 &   0.31$\pm$0.18	& $-$1      &  +1 \\
45  &  01:33:43.5 &  30:41:04.0 &  0.49$\pm$0.14	& $-$0.04$\pm$0.73      &   $-$1\\
47  &  01:33:48.3 & 30:33:05.2 &   0.96$\pm$0.18  & $-$0.48$\pm$0.63      &  $-$1 \\
53  &  01:33:54.3 & 30:33:50.5 &   0.69$\pm$0.25	& $-$0.81$\pm$0.54      &  $-$0.49$\pm$2.95 \\
54  &  01:33:54.6 & 30:45:18.9 &   0.32$\pm$0.10	& $-$1      &  \nodata \\
55  &  01:33:54.9 & 30:33:10.9 &   10.5$\pm$0.50	& $-$0.78$\pm$0.44      &  $-$0.93$\pm$0.33\\ 
60  &  01:33:58.5 & 30:36:24.6 &   0.34$\pm$0.12	& $-$0.70$\pm$0.62      &  $-$1 \\
62  &  01:33:58.5 & 30:33:33.7 &   0.90$\pm$0.26  &  $-$1      &  $-$1 \\
64  &  01:34:00.3 & 30:42:18.6 &   1.34$\pm$0.20    & $-$0.70$\pm$0.52      &  $-$0.54$\pm$0.84 \\
66  &  01:34:01.5 & 30:35:19.3 &   1.44$\pm$0.28    & $-$0.33$\pm$0.70      &  $-$0.09$\pm$0.78 \\
73  &  01:34:10.7 & 30:42:23.8 &   1.71$\pm$0.19    &  $-$0.70$\pm$0.51      &  $-$1 \\
78  &  01:34:14.3 & 30:53:54.1 &   0.47$\pm$0.11	&    $-$0.28$\pm$0.71      &  $-$0.54$\pm$0.84 \\
83  &  01:34:16.5 & 30:51:54.5 &   0.44$\pm$0.09	&   $-$0.76$\pm$0.49      &  $-$1\\ 
85  &  01:34:17.5 & 30:41:23.2 &   0.27$\pm$0.10	&   $-$1      &  +1 \\
94  &  01:34:33.0 & 30:46:38.3 &   0.41$\pm$0.07	&   $-$0.25$\pm$0.70      &  $-$1 \\
97  &  01:34:41.0 & 30:43:28.0 &   0.73$\pm$0.10	&   $-$0.65$\pm$0.55      &  $-$0.63$\pm$0.75 \\
270\tablenotemark{f} &  01:34:23.3 & 30:54:24.0 &   0.29$\pm$0.08	&   $-$0.36$\pm$0.49      &  $-$0.13$\pm$0.74
\enddata
\tablenotetext{a}{IDs from the optical SNR catalogue of GKL98.}
\tablenotetext{b}{Coordinates measured from the computed centroid of the source from the soft-band image.}
\tablenotetext{c}{Count rate in the broad band spectrum, 0.35$-$8.0 keV.}
\tablenotetext{d}{S\,=\,0.35$-$1.1 keV (soft band), M = 1.1$-$2.6 keV (medium band), H = 2.6$-$8 keV (hard band). }
\tablenotetext{e}{Hardness ratios set to $-$1 when there are zero counts in the harder band of the ratio, +1 when there are
zero counts in the softer band of the ratio.  }
\tablenotetext{f}{This is a new optical identification for an X-ray source designated as CXOM33 J013441.0+3043280 (see
Section 7 for details) and matching source 270 from the \xmm\, catalogue of Pietsch \etal\, (2004).  }
\end{deluxetable}

\begin{deluxetable}{cccc}
\tabletypesize{\footnotesize}
\tablecaption{Spectral Fits to X-ray Counterparts of Optical SNRs\tablenotemark{a}}
\tablehead{
\colhead{ID\tablenotemark{b}} & \colhead{$kT$ (keV)}  &    
\colhead{$L_{\,35}$(0.35$-$8) (ergs s$^{-1}$) }  &  \colhead{$\chi^{2}$ (d.o.f.) }  }
\startdata
9	&	0.30	&	1.9 	& 31 (43)\\
27	&	0.34	&	1.9 	& 26 (36)\\
28	&	0.37	&	31.2	& 54 (70)\\
29	&	0.30	&	 17.3	& 35 (70\\
31	&	0.34	&	 32.8	& 124 (96)\\
35	&	0.38	&	 13.4	& 27 (56)\\
37	&	0.47	&	 0.5	& 36 (46)\\
45	&	0.85	&	 0.6	& 17 (32)\\
47	&	0.30	&	 2.9	& 35 (39)\\
53	&	0.24	&	 0.9	& 43 (82)\\
54	&	0.23	&	 2.2	& 5 (14)\\
55	&	0.29	&	 43.2	& 107 (104)\\
60	&	0.18	&	 1.6 	& 18 (70)\\
62	&	0.23	&	 4.2	& 67 (70)\\
64	&	0.35	&	 1.2	& 47 (53)\\
66	&	0.28	&	 1.7	& 69 (108)\\
73	&	0.31	&	 10.7	& 29 (41)\\
78	&	0.18	&	 4.7	& 32 (65)\\
83	&	0.22	&	 4.6	& 23 (39)\\
85	&	0.21	&	 2.4	& 47 (50)\\
94	&	0.62	&	 1.4	& 20 (31)\\
97	&	0.35	&	 3.1	& 41 (52)\\
270\tablenotemark{c} &	0.33	&	2.3	&	17 (38)\\
\enddata
\tablenotetext{a}{All fits are computed assuming a Raymond-Smith model and
a distance of 795 kpc to M33.  The luminosity of SNR
55 was computed with N$_{H}$(Gal)\,=\,5.4$\times$10$^{20}$ cm$^{-2}$ (frozen), abundances 0.4 solar (frozen)
and with N$_{H}$(M33), kT and spectrum normalization as free parameters.  Spectra of the remaining remnants
in the table were computed using N$_{H}$(Gal)\,=\,5.4$\times$10$^{20}$ cm$^{-2}$ (frozen),
N$_{H}$(M33)\,=\,7$\times$10$^{20}$ cm$^{-2}$ (best fit value from SNR 55; now frozen) and with
$kT$ and spectrum normalization as free parameters.}
\tablenotetext{b}{IDs from the optical SNR catalogue of GKL98)}
\tablenotetext{c}{Emission parameters for a \chan\, source matching a newly identified optical SNR (see Section 7
for details).  The object matches source 270 from the \xmm\, catalogue of Pietsch \etal\, (2004).  }
\end{deluxetable}

\clearpage

\end{document}